\documentclass[10pt,letterpaper,twocolumn,prl,showpacs]{revtex4-1}

\usepackage[english]{babel}

\usepackage{amsmath,amssymb}
\usepackage[dvips]{graphicx,color}

\begin{document}

\title{Measurement of microresonator frequency comb stability by spectral interferometry}

\author{K. E. Webb$^*$, J. K. Jang, J. Anthony, S. Coen, M. Erkintalo, and S. G. Murdoch}

\affiliation{The Dodd-Walls Centre for Photonic and Quantum Technologies, Department of Physics, The University of Auckland, Private Bag 92019, Auckland 1142, New Zealand \\
$^*$Corresponding author: kweb034@aucklanduni.ac.nz }

\begin{abstract}
We demonstrate a new technique for the experimental measurement of the spectral coherence of microresonator optical frequency combs. Specifically, we use a spectral interference method, typically used in the context of supercontinuum generation, to explore the variation of the complex degree of first order coherence across the full comb bandwidth. We measure the coherence of two different frequency combs, and observe wholly different coherence characteristics. In particular, we find that the observed dynamical regimes are similar to the stable and unstable modulation instability regimes reported in previous theoretical studies. Results from numerical simulations are found to be in good agreement with experimental observations. In addition to demonstrating a new technique to assess comb stability, our results provide strong experimental support for previous theoretical analyses.
\end{abstract}

\maketitle

Microresonator-based frequency combs have attracted tremendous interest in recent years \cite{delhaye07,kippenberg11,delhaye11,okawachi11,ferdous11,papp11,herr12}, thanks to their numerous potential applications such as waveform synthesis \cite{ferdous11}, telecommunications \cite{pfeifle14}, and optical clocks \cite{papp14}. These combs possess a number of attractive features compared to their mode-locked laser-based counterparts, including large comb spacing and the possibility for on-chip integration \cite{kippenberg11}. They arise through third-order ``Kerr'' nonlinear optical effects: a continuous-wave laser driving a high-$Q$ microresonator gives rise to spectral sidebands via modulation instability (MI), which subsequently interact to generate new frequency components through cascaded four-wave mixing \cite{delhaye07}. In the time domain, such ``Kerr combs'' can correspond to periodic MI patterns or temporal cavity solitons \cite{leo10}, that can either be stable or exhibit fluctuations over consecutive roundtrips \cite{coen13b}.

Due to the high precision required by many frequency comb applications, the stability and noise characteristics of comb sources are of great interest. So far, the coherence of microresonator-based combs has been examined predominantly using RF beat note \cite{papp11,herr12,herr14} and optical heterodyne measurements \cite{herr12,wang13}. Unfortunately, these techniques do not easily reveal how the coherence varies across the comb bandwidth. In particular, RF measurements only yield an average over all of the comb lines, while the optical heterodyne technique delivers information on a single line at a time, therefore requiring multiple measurements to gain quantitative insights across the entire comb. Spectral line-by-line pulse shaping can give an indication of the stability of several comb lines at once \cite{ferdous11,loh14}, but that method does not provide a quantitative measure of coherence, nor is it applicable to unstable combs. An alternative technique was recently proposed to alleviate these deficiencies \cite{erkintalo14,torrescompany14}, motivated by prior investigations of optical supercontinuum coherence \cite{bellini00,dudley02,gu03,lu04,dudley06}. In \cite{erkintalo14,torrescompany14} the stability of numerically simulated Kerr frequency combs was quantified in terms of the wavelength-dependent complex degree of first-order coherence, which was shown to allow for quantitative analysis of stability across the full comb spectrum. Moreover, it was shown that different comb operating regimes, previously identified in \cite{coen13b}, are associated with unique coherence characteristics, suggesting that measurements of the wavelength-dependent degree of coherence could also permit the comb operating regime to be identified. However, to the best of our knowledge, an experimental measurement of the first order coherence of Kerr frequency combs has not yet been demonstrated.

In this Letter, we experimentally measure the wavelength-dependent degree of coherence for microresonator frequency combs, demonstrating how this approach provides direct and easy access to the stability across the entire comb bandwidth. To qualitatively confirm our coherence measurements, we also carry out optical heterodyne measurements of selected comb lines. We measure the coherence of two different comb regimes, observing very different characteristics that are in good agreement with features expected for combs in the stable and unstable MI regimes \cite{coen13b}. Numerical simulations of the comb spectra and coherence are also in good agreement with our experimental results. We expect our work to facilitate future investigations of Kerr comb stability and operating regimes.

In the context of microresonator frequency combs, the modulus of the complex degree of first order coherence describes wavelength-dependent correlations of cavity outputs separated by a fixed coherence delay $|t_1-t_2|$  \cite{erkintalo14,torrescompany14}. It is defined as
\begin{equation}
|g^{(1)}_{12}(\lambda,t_1-t_2)|=\frac{|\langle\tilde{E}^\ast(t_1,\lambda)\tilde{E}(t_2,\lambda)\rangle|}{S(\lambda)},
\label{eq:g12}
\end{equation}
where $\tilde{E}(t,\lambda)$ is the complex Fourier transform of the output field at time $t$, and $S(\lambda)$ is the mean spectrum. Provided that the fields have the same amplitude, this quantity is equal to the visibility of the spectral fringes that arise from the interference of the field with a delayed copy of itself. Accordingly, it can be measured experimentally by passing the resonator output through a delayed interferometer and measuring the visibility of the resulting spectral fringes \cite{erkintalo14}. Due to the high finesse of the resonator, useful measurements require a coherence delay of the order of one photon lifetime \cite{erkintalo14}. Moreover, the measurement should ideally be repeated for several coherence delays so as to obtain insights on the comb operation regime.

\begin{figure}
\centering
\includegraphics[width=8.5cm]{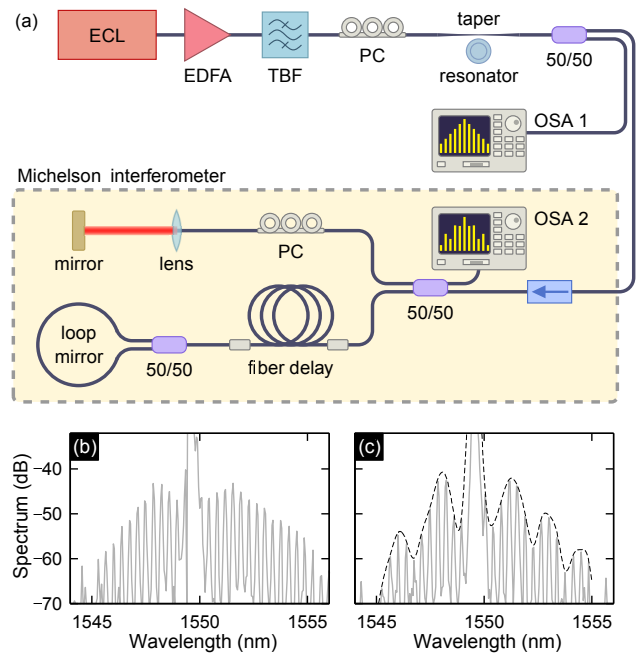}
\caption{(a)~Experimental setup. PC: polarization controller. (b)~Example of frequency comb spectrum measured on OSA~1. (c)~Example of the spectral fringes observed for the same comb as in (b), measured on OSA~2.}
\label{fig:setup}
\end{figure}

\looseness=-1 Figure~\ref{fig:setup}(a) shows the experimental setup used to measure the coherence of microresonator frequency combs. An external cavity laser (ECL) at 1549.5~nm is amplified with a C-band erbium-doped fiber amplifier (EDFA) and then filtered with a 0.4~nm tunable bandpass filter (TBF) to produce an approximately 10~mW continuous-wave pump. The pump is launched into a fiber taper with waist diameter $\approx1~\mu$m, which is coupled to a fused silica microresonator. The resonator was cut with a CO$_2$ laser using a method similar to \cite{delhaye13}. It has a major diameter \mbox{$D_\text{maj}\approx1.5$~mm}, minor diameter \mbox{$D_\text{min}\approx40~\mu$m}, finesse \mbox{$\mathcal{F}=2.3\cdot10^5$} ($Q=9.7\cdot10^8$, measured using the ringdown method), and a free spectral range \mbox{$\text{FSR}=43$~GHz}. Using a commercial finite-element solver (COMSOL Multiphysics 5.1), we calculated the resonator to have a group-velocity dispersion coefficient \mbox{$\beta_2=-21.9$~ps$^2$/km} and nonlinear interaction coefficient \mbox{$\gamma=1.5$~W$^{-1}$km$^{-1}$} at the pump wavelength. The resonator coupling is optimized by changing the pump polarization and resonator position relative to the taper. The wavelength of the ECL is tuned until it reaches a cavity resonance that produces a frequency comb, and that thermally locks itself to the pump \cite{carmon04}. Half of the taper output is sent to an optical spectrum analyzer [OSA 1 in Fig.~\ref{fig:setup}(a)] so as to measure the mean comb spectrum. A typical example of a generated comb is shown in Fig.~\ref{fig:setup}(b). The other half of the taper output is input either to a Michelson interferometer to measure the visibility across the comb, or to a delayed self-heterodyne interferometer (DSHI) for linewidth measurements \cite{derickson}.

The coherence across the entire comb spectrum is extracted from the visibility measured at the output of the Michelson interferometer. The resonator output is first passed through an optical isolator, then through a 50/50 coupler which acts as the beamsplitter in the interferometer. One arm contains a free-space retroreflecting mirror, while the second arm contains a fiber loop mirror and an interchangeable fiber segment for control of the coherence delay. The outputs of the two arms are recombined at the 50/50 coupler such that they are co-polarized and their powers are balanced. The output of the interferometer is then recorded on a second OSA [OSA 2 in Fig.~\ref{fig:setup}(a)]. Figure~\ref{fig:setup}(c) shows an example of the observed spectral fringes for the same comb as shown in Fig.~\ref{fig:setup}(b).

Since the resonator we use has a high $Q$, we require long fiber lengths to investigate correlations over delays greater than one photon lifetime ($t_\text{ph}\approx0.8~\mu$s, corresponding to approximately 170~m of silica fiber). Such long fiber delays cause the interferometer to be unstable, which prevents the interferometer phase from being systematically fine-tuned, as well as the visibility from being extracted from a single OSA scan. However, we can use the instability to our advantage by taking multiple recordings of the interferometer output spectrum. Specifically, since the interferometer phase fluctuates randomly, and is different for each OSA scan, we are able to sample through all possible phases simply by recording several spectra on the OSA for each fiber delay length. From these realizations we extract the maximum and minimum power level of each comb line, which allows us to evaluate the corresponding interference visibility. To ensure that possible comb power fluctuations do not influence the measurement, we simultaneously record the microresonator output outside of the interferometer using the first OSA (OSA 1 in Fig.~\ref{fig:setup}). For the results that follow, the visibility was extracted from a set of 50 OSA measured interference patterns.

\begin{figure}
\centering
\includegraphics[width=\linewidth,clip=true]{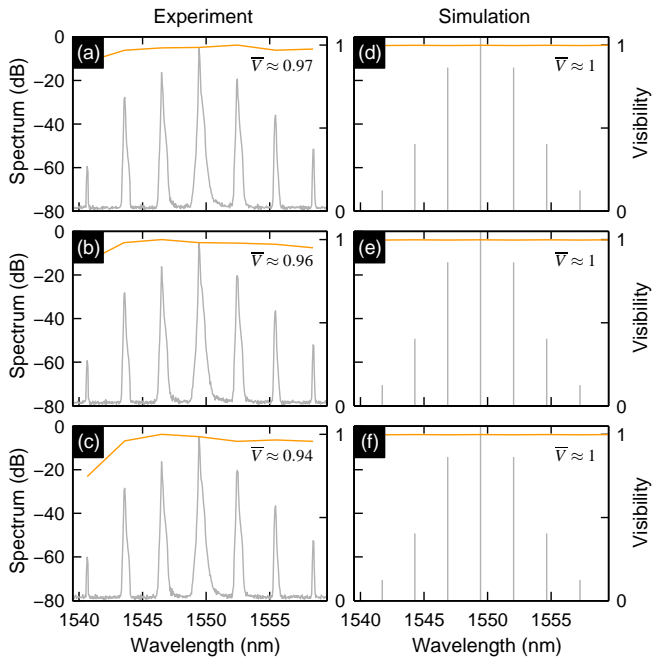}
\caption{Orange solid lines show (a)--(c) experimentally measured and (d)--(f) numerically simulated comb visibilities of an 8~FSR comb at interferometer delays (a,d) 0~m, (b,e) 50~m, and (c,f) 200~m (right axis). Gray lines show measured and simulated comb spectra for reference (left axis). Simulations use detuning $\Delta=-10.7$. Average visibilities $\overline{V}$ are shown in top right corner of each subfigure.}
\label{fig:smi}
\end{figure}

\begin{figure}
\centering
\includegraphics[width=\columnwidth]{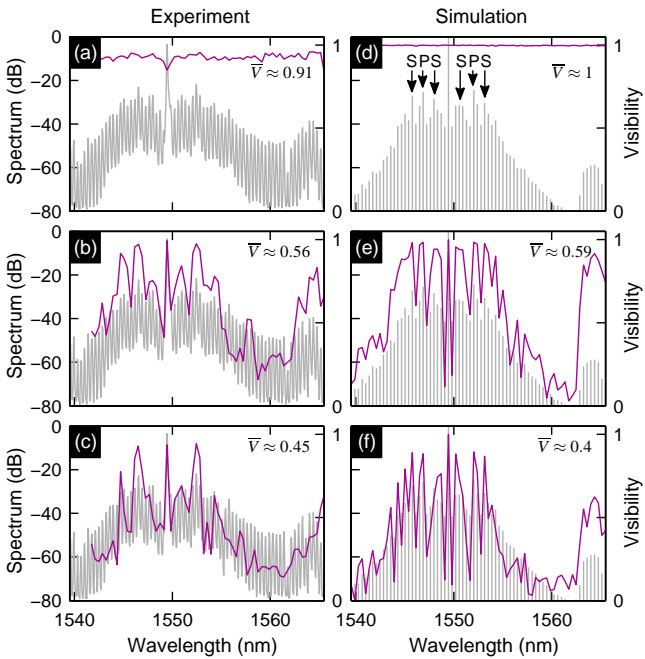}
\caption{Purple solid lines show (a)--(c) experimentally measured and (d)--(f) numerically simulated comb visibilities of a 1~FSR comb at interferometer delays (a,d) 0~m, (b,e) 50~m, and (c,f) 200~m (right axis). Gray lines show measured and simulated comb spectra for reference (left axis). Simulations use detuning $\Delta=-8$. Average visibilities $\overline{V}$ are shown in top right corner of each subfigure.}
\label{fig:umi}
\end{figure}

In addition to experimental measurements, we also carry out numerical simulations to further provide support for our results, and also to gain more insight into the comb characteristics. Our simulations use the generalized Lugiato-Lefever equation \cite{coen13b,erkintalo14,coen13a}, which we integrate using the split-step Fourier method, with parameters corresponding to our experiment as given above. Using the normalization in \cite{leo10,coen13b}, these parameters correspond to a driving power of $X = 360$. The normalized cavity detuning $\Delta\approx\mathcal{F}\cdot(\omega_0-\omega_\mathrm{p})/(\pi\cdot\mathrm{FSR})$, where $\omega_0$ and $\omega_\mathrm{p}$ are the frequencies of the pumped mode and driving laser, respectively, is chosen for each comb to best match the experimental results (see captions of Figs.~\ref{fig:smi} and \ref{fig:umi}).

\looseness=-1 Figures~\ref{fig:smi}(a)--(c) show the experimentally measured spectral visibilities at fiber delays of 0~m~($0\cdot t_\text{ph}$), 50~m~($0.3\cdot t_\text{ph}$), and 200~m~($1.2\cdot t_\text{ph}$), respectively, for a frequency comb which has a mode spacing of 8 FSR. (For clarity, in each panel the average comb spectrum is superimposed with the visibilities.) Also shown for each delay is the average visibility $\overline{V}$, taken over all comb modes above $-60$~dB (relative to the pump). Corresponding numerical simulations are shown in \mbox{Figs.~\ref{fig:smi}(d)--(f)}, and they are in good agreement with experimental results (discrepancies are attributed to imperfect knowledge of resonator parameters, in particular the cavity detuning). We find that, for all coherence delays, all of the spectral components lead to high visibility fringes, indicating a state of high coherence. The comb line at 1540~nm displays some degradation in visibility for increasing fiber delay; this is most likely due to poor signal-to-noise ratio at this wavelength.

By tuning the laser wavelength to a different resonance, we access another comb regime, with spectrum and spectral visibility shown in Figs.~\ref{fig:umi}(a)--(c). The corresponding simulation results are shown in Figs.~\ref{fig:umi}(d)--(f), and they are again in good agreement with our experiments. The average visibility for each delay is also shown, and there is good agreement between experiments and simulations. This comb has a dispersive-wave-like feature around 1564~nm \cite{coen13a}, which we believe is due to an avoided mode crossing. We include the mode crossing in our simulations by introducing an additional phase shift to the modes in the vicinity of the crossing \cite{herr14b,xue15}. 

At zero delay, the comb again shows a high visibility across the spectrum ($\overline{V}\approx0.91$). At a delay of 50~m, however, only the pump and selected lines on both sides of the pump maintain high visibility, whilst the average visibility across the comb is reduced to $\overline{V}\approx0.56$. In fact, a more detailed examination of our numerical simulations shows that the lines with elevated coherence correspond to the primary and secondary comb components that emerge in the initial stages of comb formation, as described in \cite{herr12}. (For clarity, the primary and secondary lines are labelled in Fig.~\ref{fig:umi}(d) as P and S, respectively.) Also the modes around 1564~nm display enhanced visibility, which is captured by the numerical simulations. However, further increasing the delay to 200~m reveals that only the pump and the primary MI sidebands are highly coherent, while the rest of the spectrum has a greatly reduced visibility ($\overline{V}\approx0.45$). Although not shown here, other results taken over a range of fiber delays are also in good agreement with the simulations.

Our measurement technique provides insight on the stability of each comb line simultaneously, which represents a clear advantage to conventional methods. To confirm that our coherence measurements agree with more conventional linewidth measurements, we used a DSHI \cite{derickson} to measure the linewidths of selected comb lines representative of the comb dynamics seen in Fig.~\ref{fig:umi}. We first amplify the comb using a C-band EDFA, then filter out individual comb lines with a 0.4~nm TBF. The filter output is then passed through a 50/50 coupler, where one of the outputs is launched into approximately 30~km of standard telecommunications single-mode fiber to ensure that the fields are uncorrelated, while the other is phase modulated at $80$~MHz. The two arms of the interferometer are recombined with a second 50/50 coupler so that the fields are co-polarized and of equal power. The $80$~MHz beat note is recorded on an electrical spectrum analyzer. Assuming a Gaussian lineshape, we obtain a linewidth of 61~kHz for our pump. In line with the high-degree of measured coherence, the primary comb line (8~FSR from the pump) has a linewidth very close to that of the pump (64~kHz), while the secondary line (4~FSR from the pump) is slightly broadened to 89~kHz, indicating a lower coherence compared to the pump. We also measured a line from the subcomb surrounding the primary comb line (7~FSR from the pump), and obtained a linewidth of 386~kHz, which is significantly broadened compared to the pump linewidth. These results qualitatively agree with the spectral coherence measurements presented in Fig.~\ref{fig:umi}.

\begin{figure}
\centering
\includegraphics[clip=true,width=\linewidth]{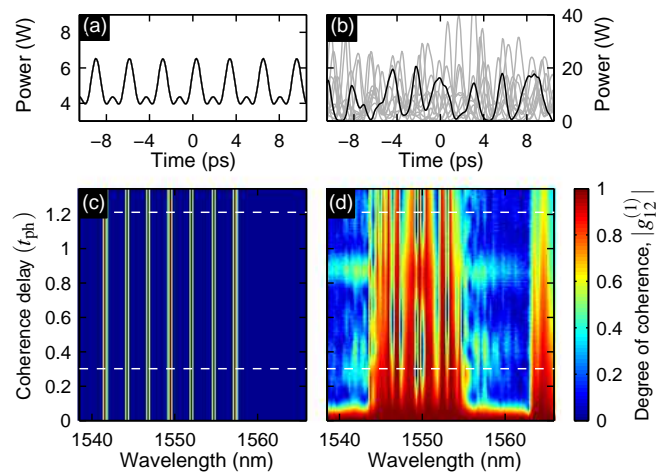}
\caption{(a,b) Snapshots of the numerically simulated temporal intracavity field, taken at 40000 roundtrip ($1.1\cdot t_\text{ph}$) intervals (gray lines). Black line shows a particular realization in more detail. (c,d) Corresponding simulated spectral coherence as a function of coherence delay. Dashed lines in (c,d) indicate the simulation results previously shown in Figs.~\ref{fig:smi} and \ref{fig:umi}.}
\label{fig:cohsim}
\end{figure}

The comb characteristics shown in Figs.~\ref{fig:smi} and \ref{fig:umi} are reminiscent, respectively, of the stable and unstable MI regimes identified in previous studies \cite{erkintalo14,coen13b}. Closer analysis of our numerical simulations confirm this hypothesis. Figures~\ref{fig:cohsim}(a)~and~(b) show temporal snapshots of the simulated intracavity fields corresponding to the combs shown in Figs.~\ref{fig:smi} and \ref{fig:umi}, respectively. The profile in Fig.~\ref{fig:cohsim}(a) shows a fully periodic and stable MI pattern, whilst significant fluctuations can be observed for the profile in Fig.~\ref{fig:cohsim}(b). In Figs.~\ref{fig:cohsim}(c) and (d), we show the corresponding numerically simulated degree of spectral coherence for both combs, evaluated for a wide range of coherence delays (the dashed lines correspond to the results previously shown in Figs.~\ref{fig:smi} and \ref{fig:umi}). We see that all of the spectral modes of the stable patterned field exhibit perfect visibility for all delays [see Fig.~\ref{fig:cohsim}(c)]. In contrast, the spectral visibility associated with the field shown in Fig.~\ref{fig:cohsim}(b) degrades quickly for almost all comb modes; it is only the primary and secondary lines that have high coherence. It is clear from Figs.~\ref{fig:cohsim}(a,c) and (b,d) that the simulated combs in Figs.~\ref{fig:smi} and \ref{fig:umi} are operating in the stable and unstable MI regimes \cite{erkintalo14}, respectively. Moreover, although not shown here, more extensive simulations predict that combs operating in the cavity soliton regimes would yield coherence characteristics \cite{erkintalo14} incompatible with our experimental observations. We can thus conclude with confidence that the experimental results shown in Figs.~\ref{fig:smi} and \ref{fig:umi} are indicative of combs operating in the stable and unstable MI regimes, respectively.

To summarize, we have experimentally and numerically explored the complex degree of first-order coherence of microresonator-based frequency combs across the entire comb bandwidth. We have experimentally measured the coherence for two different regimes, and found their characteristics to agree very well with those previously predicted for combs in the stable and unstable MI regimes. Moreover, our results show quantitative differences between the two operation regimes, highlighting how this technique can be used to distinguish between different Kerr comb operating regimes.  We expect our method to have wide impact in future studies, enabling the stability and operating regime of microresonator optical frequency combs to be easily determined.

\looseness=-1 We acknowledge support from the Marsden Fund of The Royal Society of New Zealand. M. Erkintalo also acknowledges support from the Finnish Cultural Foundation.


\newcommand{\enquote}[1]{``#1''}

\end{document}